\documentclass{icrc2009}

\usepackage{graphicx}   
\usepackage{caption}    
\usepackage[font=footnotesize]{subfig} 
\usepackage{fixltx2e}
\usepackage{url}

\newcommand{\shorttitle}[1]%
{\markboth{Proceedings of the 31\MakeLowercase{$^{st}$} ICRC, {\L}\'{o}d\'{z} }{#1} }
\newcommand{\etal}{\MakeLowercase{\textit{et al. }}} 

\begin{document}
\title{Rigidity dependence of  cosmic ray
   escape length in the Galaxy obtained  from
  a comparison of proton and iron spectra  in  the range 3-3000 GV}

\author{\IEEEauthorblockN{Olga Strelnikova\IEEEauthorrefmark{1},
			  Lyubov Sveshnikova\IEEEauthorrefmark{1} and\\
                          Vladimir Ptuskin\IEEEauthorrefmark{2}}
                            \\
\IEEEauthorblockA{\IEEEauthorrefmark{1}Skobeltsyn Institute of Nuclear Physics of Moscow State University,\\ Leninskie Gory, MSU, Moscow 119992, Russia}
\IEEEauthorblockA{\IEEEauthorrefmark{2}Pushkov Institute of Terrestrial Magnetism, Ionosphere and Radio Wave Propagation (IZMIRAN),\\ Russian Academy of Sciences, Troitsk, Moscow Region 142190, Russia}}

\shorttitle{Strelnikova \etal Rigidity dependence of  cosmic ray
   escape length}
\maketitle

\begin{abstract}
 The simple leaky-box
 model of propagation of cosmic rays in the Galaxy is quite suitable
 for handling  data on cosmic ray  nuclei
 energy spectra and composition at $E \gg 1$ GeV [1,2]. In the leaky-box
 model a full information about cosmic  ray propagation in Galaxy
 is  compressed to the single parameter - escape length,
 $X_{e}$, characterizing mean grams of a matter passed by cosmic
 rays from sources to the Earth. In this paper we analyze the 
 world data on proton and iron cosmic ray spectra collected
 in the past (HEAO, CRN et al.) and in series of recent electronic
 experiments (ATIC, CREAM, AMS, BESS, Tracer et al.) and obtain
 the rigidity dependence of escape length, 
  $X_{e}(R)=\sim R^{-0.47\pm-0.03}$,  from the measured 
  rigidity dependence of the 
  protons/iron ratio. It  quite agrees 
 with the one estimated in standard manner from the secondary/primary
 nuclei ratio. But at $R > 300$ GV the behavior of $X_{e}(R)$ 
  distinctly changes, that can  (variant of  explanation)
  point out to the change of proton/iron 
 ratio in cosmic  ray sources. 

\end{abstract}

 
\section{Introduction}
 In spite of apparent absence of physical background for
 the leaky-box model, where  the transport of energetic particles
 is described by introducing
the mean escape time of cosmic rays from the Galaxy, and 
the cosmic-ray density, the source density,
 the gas density  do not depend on coordinates, it may be applied 
to the 
 study of diffusion and nuclear
spallation of stable nuclei
attended by the production of secondary relativistic nuclei in
the interstellar gas \cite{Berezinskii} \cite{Main}. This
can be explained by the concentration of cosmic rays sources
and the interstellar gas in a relatively thin galactic disk immersed
in the flat but thick cosmic-ray halo \cite{Berezinskii}.
The spatial distribution in the low-density halo
is the same for different stable nuclei because of the negligible nuclear
spallation. The calculated intensities of stable nuclei for an
observer at the galactic disk look as corresponding leaky-box
expressions even for nuclei with large cross sections and all propagation
is described by some escape length of cosmic rays from the
Galaxy $X_e$ (measured in g/cm$^2$) that is a function of particle rigidity $R$.
This  important  parameter means grams of a matter passed by cosmic
 rays from sources to the Earth in average.
 In  \cite{Main}    
 there was  proposed a  way  to estimate the analogous
 parameter  in the diffusion
model - effective values of $X_{ef}$. It was shown that  leaky-box 
 is a good approximation
 to the widely known and used for the  interpretation of various
  cosmic-ray data  basic GALPROP (Galactic Propagation) model
 \cite{GALPROP} with $1\% $  accuracy for all nuclei. In \cite{Main}
values of $X_{ef}$  were found for the three set of GALPROP parameters,
corresponding to the three models of the propagations.
\noindent
 1) Plain diffusion
model (PD) 

$$ X_{ef}=19 \beta ^3~ \rm{g/cm}^2 ~R < 3~\rm{GV},$$
$$  X_{ef}=19\beta^3(R/3~\rm{GV})^{-0.6} \rm{g/cm}^2,   R >  3~\rm{GV}; \eqno(1)$$ 
2)Diffision with  acceleration (DR) 

$$ X_{ef}=7.2(R/~\rm{3GV})^{-0.34}\rm{g/cm}^2,R >  30~\rm{GV};\eqno(2)$$ 
3)Diffusion with damping (DRD)\\
$$ X_{ef}=13 (R/3~\rm{GV})^{-0.5}\rm{g/cm}^2,~R >  10~\rm{GV}\eqno(3)$$   

For the comparison we present also the widely used approximation   \cite{Jones}:

$$ X_{ef}=11.8 \beta(R/4.9~\rm{GV})^{-0.54} \rm{g/cm^2} ,R > 4.9~\rm{GV}\eqno(4)$$

   \begin{table*}[th]
  \caption{
  The fit of the ratio of $I_H/I_{Fe}-\rm{fit}$  by the polynomial function
  of the fifth  order. Averaged points of the ratio $I_H/I_{Fe}-\rm{points}$ are calculated
  in  14
  R-bins
  with the corresponding errors
}
  \label{table1}
  \centering
  \begin{tabular}{|c|c|c|c|c|c|c|c|c|c|c|c|c|c|}
  \hline
R, GV   & 3.16&  5  &  8   &  12.6&   20 &  31.6&   50 &    89 &  178&   355&    708&   1412&  2818  \\
$Ip/Ife-\rm{fit}$& 2907& 2313&  1875&  1564&  1347&  1197&  1095&  1012 &  952&   914 &    889&   877 &  894  \\ 
Errors      & 249 & 212 &  105 &   47 &   43 &   101&   41 &    56 &   65&    84 &    143&   166 &  150  \\ 

$Ip/Ife-\rm{points}$& 2432& 2281&  1748&  1676&  1235&  1203&  1067&   991 &  842&   1053&    955&   976 &  1274 \\ 
Errors  & 249 & 212 &  105 &   50 &   40 &   101&   40 &    54 &   57&    97 &    154&   185 &  447  \\ 

  \hline
  \end{tabular}
  \end{table*}

 The  leaky-box approach due to its simplicity until very recent was  used
by many authors.

Intensity $I_{A}$ of CR nuclei  with mass number $A$ near the Earth
is connected with their spectrum in a source $Q_{A}$(if neglecting  the energy losses at small
energies and the contribution from the fragments produced from heavier nuclei)
by the simple equation:

$$I_A=\frac{Q_{A}(R)}{4 \pi \rho}\times
\frac{1}{ \frac{1}{X_{ef}}+\frac{1}{X_{in}^A}  },\eqno(5)$$

\noindent
where $\rho$ - gas density, $X_{in}^A$ - interaction length for the nucleus with mass number A.
(calculated from GALPROP). For protons $X_{in}^H$ $\sim$   110 g/cm$^2$ at 1 GeV and it decreases to  70 g/cm$^2$ 
at  10000 GV,  that is $X_{ef}\ll X_{in}^A$, from whence the well  known
equation 
arises: $$I_H\sim Q_H*X_{ef},~~~   \gamma_{obs}=\gamma_{sour}+\alpha,$$
where $\alpha$  is the slope of $X_{ef}$ dependence,
$\gamma_{obs}$, $\gamma_{sour}$ - the slopes of observed and source
spectra correspondingly in the case of power-like laws.

 But while  interaction length for the iron nuclei is much smaller
$X_{in}^{Fe} \sim 2.7~\rm{g/cm}^2$, there is the rigidity region $R_{min} \div R_{max}$
 for Fe nuclei  where\\
$$I_{Fe}\sim Q_{Fe}*X_{in}^{Fe},~~~ \gamma_{obs} \sim \gamma_{sour},$$
 reaching the asymptotic value (proton and  Fe spectra are parallel),
$\gamma_{obs}=\gamma_{sour}+\alpha$ when $X_{ef}$ at $R_{max}$ becomes much smaller than
$X_{in}^{Fe}$ .


We see that one  can try  to estimate the rigidity
 dependence of  $X_{ef}$ from the ratio
of any two spectra in the interval $R_{min} \div R_{max}$, but 
 with  essential  reservations:\\
 - to bear in mind that the intrinsic
 property of the leaky-box model
is the independence of $X_e$ on cross section and as a result on
 the type of nucleus;\\
 - to assume  that all types of nuclei are produced in the same  type of sources
and the chemical composition of accelerated particles in the sources does not
depend on rigidity in the investigated interval - $$Q_{Fe}(R)/Q_{H}(R)=\rm{const};$$
 - to use  nuclei  with   significantly   different interaction lengths, besides, the contribution of fragments
among them should be  negligible; \\
 - to find way  to estimate $Q_{Fe}(R)/Q_{H}(R)$. 

Proton and iron spectra are most suitable ones for this task.

The main question is:  if available in our days experimental data
 are enough for this task?
 
\section{Experimental data}

 As we are going to analyze spectra at fixed rigidity,
 we need the spectral data measured at  remote 
 intervals of energy per particle. Interval  pointed out in title, 3-3000 GV,
 corresponds to the energy per particle interval 2.3 GeV - 3 TeV for protons  and  44 GeV - 82 TeV for iron nuclei.
 So we are doomed  to use data from different experiments. 
 Moreover only  with 
 appearance of  new  data  on iron nuclei from the experiment  Tracer \cite{TRACER2}, 
 measured in  uniquely  wide
 energy range ($E_{part}$=30 GeV-80 TeV),   this task probably can be solved.
  Data obtained in ATIC2 experiment \cite{ATIC02} 
  are very
 important  also, because  they  fill up the gap between data obtained by magnetic spectrometers
 and ones obtained with calorimeters and emulsions.

We include in the consideration 
experimental points, which  satisfy the conditions:

 \noindent
 a) energy interval  lies between 3 and 10000 GV,\\ 
 b) errors don't exceed $ 30 \%$,\\
 c) calibration on accelerators  is done (that is why Sokol experiment
   was excluded).

 Fig. 1 and Fig. 2 represent proton and Fe
spectra measured in different experiments. The points satisfying above enumerated
conditions are denoted by filled symbols.

 Proton data are got from the following experiments: MASS91 \cite{MASS}, BESS-98 \cite{BESS98}, \cite{BESS02}BESS02,
 AMS \cite{AMS1,AMS2}, \cite{Caprice98}, ATIC02 \cite{ATIC02},
 CREAM \cite{CREAM}, JACEE \cite{JACEE1,JACEE2}, RUNJOB \cite{RUNJOB}.
 Fe nuclei are from: HEAO-3 \cite{HEAO-3}, CRN \cite{CRN},  TRACER \cite{TRACER1, TRACER2},
 \cite{ATIC02},
 Sanriku \cite{Sanriku}.
 Data were reduced to fixed modulation potential $\Phi=800$ MV by
 calculation similar to  \cite{Caprice98}.

 Fortunately  points satisfying above mentioned conditions are little scattered.
 For  further work and   convenience one  needs to approximated these
 dependencies by any way.
 In the Fig. 1 and 2  we denote by thick line the best fit of
 corresponding 
 scatter points, that is a polynomial function  of the fifth  order.
  In the Table 1  this  fit is denoted as " $I_H/I_{Fe}-\rm{fit}$ " 
 with errors
 in the upper two rows. 

 Moreover,
 we calculated averaged points in the every from 14 R-bins getting  errors
 which
 include  statistical significance and  
  a divergence
 of experimental points fallen in the fixed bin
 (thin lines with errors in Fig. 1,2.). In the Table 1
  averaged points are denoted as " $I_H/I_{Fe}-\rm{points} $ " and they are 
presented in  the lower  two rows.

 \begin{figure*}[t]
  \centering
  \includegraphics[width=5in]{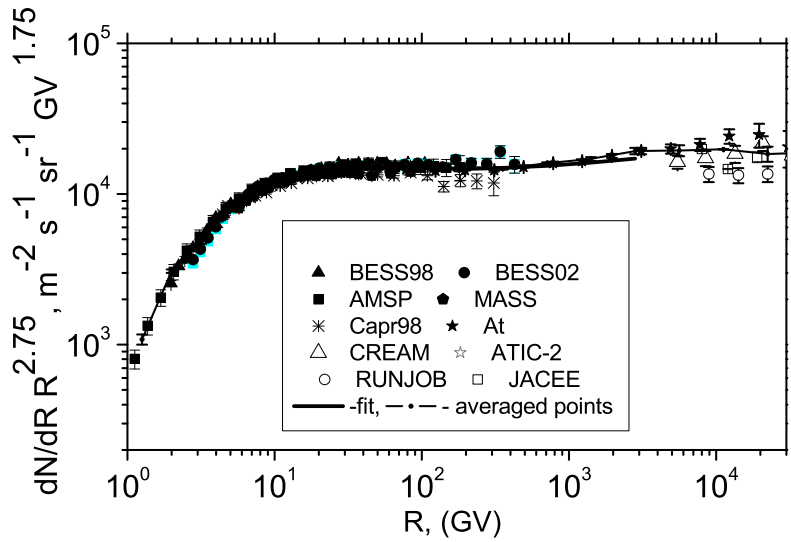}
  \caption{Proton  spectra measured in different
 experiments (for ref. see text), reduced to the modulation potential $\Phi=800$ MV; 
 open   symbols -  points with more than  $30\%$; thick line - the 5th order 
 polynomial fit, thin lines - averaged points}
  \label{fig3}
 \end{figure*}

 \begin{figure*}[t]
  \centering
  \includegraphics[width=5in]{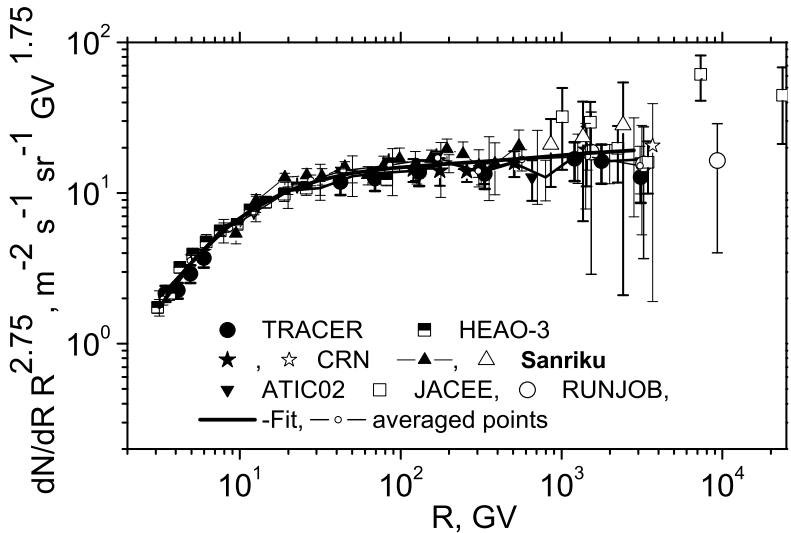}
  \caption{Spectra of iron nuclei; open   symbols -  points with more than  $30\%$; thick line - the 5th order 
 polynomial fit, thin lines - averaged points  
}
  \label{fig12}
 \end{figure*}

 \section{Results and discussion}

  The leaky-box Ed. (5) written for the proton  and Fe spectra 
   allows us to  express the value of effective escape length $X_{ef}$ 
  by means of the ratio of proton/iron measured spectra 
  $I_{H/Fe}=I_H/I_{Fe}(R)$ (from Table 1),  
  the   source chemical composition  $Q_{H/Fe}=Q_H/Q_{Fe}$ 
  (which does not
depend on rigidity and should be determined any how)
and known values of
  $X_{in}^{Fe},~~ X_{in}^{H}$:

  $$X_{ef}=\frac{I_{H/Fe}(R)/Q_{H/Fe}-1}
  {
\frac{1}{X_{in}^{Fe}}  +  \frac{I_{H/Fe}(R)/Q_{H/Fe}} {X_{in}^H}}.\eqno(6)$$

  The Eq. (6) will work obviously only
  in the range of rigidities $R_{min}-R_{max}$
  where   $X_{ef}$ is not much smaller than  $X_{in}^{Fe}$: at  
   $X_{ef} \rightarrow 0$ $I_{H/Fe}(R)\rightarrow Q_{H/Fe}$.
To demonstrate expected application range 
  of Eq. (6) we choose
  the critical value $R_{max}$ corresponding 
   to   $X_{ef} = 0.1 X_{in}^{Fe}~\sim 0.27 \rm{g/cm}^2$. It means that 
the region $X_{ef}< 0.27 \rm{g/cm}^2$ and $R~>~R_{max}$ is the asymptotic 
 region where
spectra become parallel and not sensitive to the $X_{ef}$.
For  three   models embedded in GALPROP
 (see Introduction) the corresponding values of $R_{max}$ look like  below:\\
\noindent
1) $R_{max}=$  3.6  TV for PD model($\alpha=-0.6$),\\  
2) $R_{max}=$  6.9  TV for DRD  model($\alpha=-0.5$),\\  
3) $R_{max}=$  46.8 TV for DR  model($\alpha=-0.34$),\\     
4) $R_{max}=$ 69000 TV for RD  model($\alpha=-0.2)$ for the very flat energy
dependence  $X_{ef}=10 R^{-0.2}$ considered in \cite{Timohin} being chosen
for the explanation of the "knee" in PCR by the change of propagation mechanism.

So $R_{max}$ comparable with the experimental maximal $R=~2.8~ TV$  (see Table 1)
 may be only
for  very steep $X_{ef}$ dependencies, as in
the  cases  of PD  or DRD models. In this case   we  determine the value
of $Q_{H/Fe}$ from the asymptotic ratio of $I_{H/Fe}(R)$ taking into
 account minor corrections. We  estimate $Q_{H/Fe}=800-1000$.

 The variant of calculation by  
the Eq. (6) for the experimental values of $I_{H/Fe}(R)$ from the Table 1
and  $Q_{H/Fe}$=800 give the next approximation:
$X_{ef}(R) \sim 4.6*(R/5~\rm{GV})^{-0.65}$, that is 
  much lower than the expected dependence $X_{ef}$ estimated
by the $B/C$ ratio  (1)-(4).  This means that our
assumption of steep $X_{ef}$ decrease  with energy
R$>$ 3 TV  (as $R^{-0.5\div -0.6}$ )
is not fully correct. 

For the analysis of the set of flat dependencies $X_{ef}(R)$
 where $R_{max} \gg 3$ TV  and there is no possibility to
 estimate the asymptotic values $I_{H/Fe}(R)$
it is  proposed 
to "normalize"  $X_{ef}$ at $R=50$ GV by the one obtained from B/C
measurements.   The values of  $X_{ef}^{B/C}$
obtained from B/C - ratio occur between
2.8 g/cm$^2$ (DR model - (2)) and 3.45 g/cm$^2$ (PD model
(1)).
The point $R=$ 50 GV 
was chosen because  the large body of data on $B/C$
are in a good coincidence in this point \cite{Propagation}, 
from the other side 
at the $ R = 50 $  GV the contribution of reacceleration processes 
surely should be  small. 

Then $Q_{H/Fe}$ value could be determined using $X_{ef}^{B/C}(50 \rm{GV})$
$\approx$  2.8 -3.45 g/cm$^2$, and $I_{H/Fe}(R)(50 \rm{GV})$=
=1067+40 (from  Tabl. 1). By Eq. (6) we get the
corresponding interval for $Q_{H/Fe}$ = 490 - 535.
Substituting obtained  limit values of $Q_{H/Fe}$ and measured 
regidity dependence of proton/iron spectra from Tabl. 1 in Eq. (6) we
calculate two  variants of rigidity
dependencies of $X_{ef} (R)$, presented in Fig. 3. 
The upper one (with $Q_{H/Fe}=$490) could be approximated 
in the interval $R=3-300$ GV as\\ 
$X_{ef}(R) \sim 3.46*(R/5~\rm{GV})^{-0.47\pm 0.03}. $.
The lower one (with $Q_{H/Fe}=$535) could be approximated  as\\ 
$X_{ef}(R) \sim 2.7*(R/5~\rm{GV})^{-0.50 \pm 0.03}. $

  \begin{figure}[!th]
  \centering
  \includegraphics[width=2.5in]{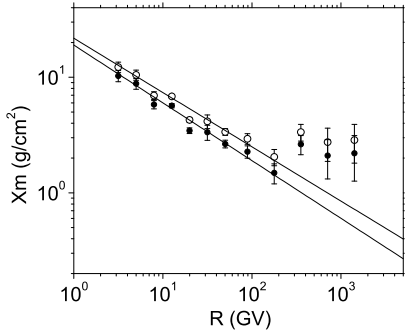}
  \caption{$X_{ef}(R)$ dependencies calculated from the measured
  proton/iron spectra (lowest two rows in Tab. 1) for the  $Q_{H/Fe}$=535
  (black circles), $Q_{H/Fe}$=490 (open circles). Thin lines
   - the approximations by  power law.}
  \label{fig3}
 \end{figure}

 Up to the $R \sim$ 300 GV 
 the obtained from protons/iron ratio rigidity dependence
 of escape length $X_{ef}(R)$ 
    can be approximated
  by the power-like law: $X_{ef}(R) \sim R^{-0.47\pm-0.03}$,
 that is in a good agreement
 with the one estimated in standard manner from secondary/primary
 nuclei ratio (1)-(4). But at $R > 300$ GV the behavior of $X_{ef}(R)$ 
  changes significantly. It is caused by  the 
  "improper" behavior of the  slopes of proton and iron spectra at 
  $R >100$ GV,  $\gamma _{Fe}$=
 2.71$\pm 0.03$ \cite{TRACER2} and $\gamma _H$=2.63 $\pm$ 0.03 \cite{ATIC02}.
  They are in  a striking contradiction with 
  the expected values: the Fe spectrum should be flatter than proton spectrum
  (see Eq.(5)). 

 Here it is worth noting that    
  in the region 300 GV-3 TV 
  (Fig.1, 2) the principal  contribution for protons
  comes from ATIC2  data \cite{ATIC02} and for  iron nuclei 
  it comes   from the Tracer  data \cite{TRACER2}.

 So if we believe that both these important  experiments
 are reliable, we should  inevitably conclude   that
  in the range $R >$ 300 GV  there is a  change of proton/iron 
 ratio in cosmic  ray sources.  Some authors of ATIC 
have come to this conclusion already
 \cite{zatsepin}.
  
  In summary we would like to stress  that the region 300 GV - 10 TV
  continues to be of importance, and the result that  spectrum
  of iron nuclei  is flatter than proton spectrum needs to be confirmed in other experiments. 
 
  {\bf The work is supported by RFBR grant  10-02-01443-a, V.S.Ptuskin -
   RFBR grant 10-02-00110a. }

\end{document}